\documentclass[superscriptaddress,aps,twocolumn,showpacs]{revtex4} 
\usepackage{epsfig}
\usepackage{graphicx}
 
\newcommand{\E}{{\rm e}} 
\newcommand{\D}{{\rm d}} 
\newcommand{\I}{{\rm i}}

\renewcommand{\Im}{{\rm Im}}

\begin{document} 
\title{Coherent transport in disordered metals: zero dimensional limit} 
\date{\today} 
\author{P. Schwab}
\affiliation{Institut f\"ur Physik, Universit\"at Augsburg, D-86135 Augsburg}
\author{R. Raimondi}
\affiliation{NEST-INFM e Dipartimento di Fisica "E. Amaldi", 
Universit\`a di Roma Tre, Via della Vasca Navale 84, 00146 Roma, Italy} 
\begin{abstract} 
We consider non-equilibrium transport in disordered conductors. 
We calculate the interaction correction to the current for a short wire connected
to electron reservoirs by resistive interfaces.
In the absence of charging effects we find a universal current-voltage-characteristics.
The relevance of our calculation for existing experiments is discussed as well
as the connection with alternative theoretical approaches.
\end{abstract} 
\maketitle 

In recent years considerable attention has been devoted to the effects of the Coulomb interaction on the
transport properties of  small structures, 
like thin diffusive films and wires \cite{altshuler85,lee85,finkelstein90,belitz94},
tunnel junctions \cite{altshuler84,schon90,ingold92}, and quantum dots \cite{aleiner01}.
One interesting issue concerns the way an applied bias voltage
affects the interaction corrections to the electrical conductivity. In diffusive metals
these corrections arise from the combination of the electron-electron 
and impurity scattering and yield well known singularities at low temperature \cite{altshuler1980}. 
It has been shown that a finite voltage
or, more in general, a non-equilibrium situation leads to a suppression of
these singularities \cite{nagaev94,raimondi1999,leadbeater2000,gutman01,schwab01}. 
In particular, in \cite{nagaev94,schwab01} non-equilibrium
transport in a short wire connected to electrical reservoirs by ideal interfaces has been considered.
However, in actual experiments the interfaces need not be ideal. 
Recently Weber et al.~\cite{weber01} investigated experimentally the non-equilibrium
transport through a metallic nano-scale bridge.
Both in \cite{weber01} and in \cite{golubev2001} it has been suggested that the 
Coulomb interaction effects are responsible
for the observed temperature dependence of the conductance and the current-voltage-characteristics.
Whereas \cite{weber01} found an agreement between theory and experiment starting from a tunneling Hamiltonian,
\cite{golubev2001} pointed out that the experimental data agree with what they expect for a diffusive conductor.
In this paper we develop a formalism in which both the resistive behavior due to the interfaces
and due to the diffusive wire region are treated on the same footing.
From our results we conclude 
that the main resistive behavior in \cite{weber01} occurs at the interfaces.

To begin with we recall the classical description of
electrical transport through structures consisting
of both interface barriers and diffusive regions. To be
definite we consider a system made by a diffusive wire of
length $L$ which is
attached to the reservoirs by two interface barriers.
We study the system in a non-equilibrium
situation with an applied voltage $V_l-V_r=V$ where the subscripts
$l$ and $r$ indicate the left and right reservoirs, respectively.
The classical resistance of the structure is the  sum
of the wire resistance and the interface resistances
$R_{\rm tot}=R_{\rm wire}+R_{l}+ R_r$,
so that the current as a function of voltage is $I=V/R_{\rm tot}$.
The microscopic calculations are conveniently carried out by using the
Keldysh formalism \cite{rammer1986}. In a disordered system  the 
Keldysh component of the Green function reads as
\begin{eqnarray}
  G^K_{\epsilon}({\bf x}, {\bf x})  &=& F_\epsilon({\bf x})
 [G^R_\epsilon({\bf x}, {\bf x})-G^A_\epsilon({\bf x}, {\bf x})] \\
& \approx & -2\pi \I {N_0} F_\epsilon({\bf x})
.\end{eqnarray}
The first line is an exact relation and defines the distribution function $F$. 
In the second line it is assumed that the density of states is a position independent constant.
The current flowing in the wire or through the boundaries is then given by
\begin{eqnarray}
I_{\rm wire} &= &e D N_0 {\cal A} \int \D \epsilon \partial_x F_\epsilon(x)
\\
I_l  & = & e \Gamma_l {\cal A} N_0 \int \D \epsilon [ F_\epsilon(0) - F_\epsilon^l] \\
I_r  & = & e \Gamma_r {\cal A} N_0 \int \D \epsilon [ F_\epsilon^r - F_\epsilon(L) ]
,\end{eqnarray}
where $D$ is the diffusion constant, ${\cal A}$ the cross section, $\Gamma_{l,r}$ are the interface
transparencies, $x= 0 \dots L$ is the position along the wire.
The reservoirs are assumed to be in thermal equilibrium, with the distribution function given by 
$F^{l,r}_\epsilon =\tanh \left[(\epsilon -eV_{l,r})/2T \right]$.
The boundary conditions
\begin{eqnarray} 
\label{eq6}
 D \partial_x F|_{x=0} &= & - \Gamma_l [ F^l - F(0) ]  \\
\label{eq7}
 D \partial_x F|_{x=L} &= & - \Gamma_r [ F(L)- F^r  ]
\end{eqnarray}
guarantee current conservation at the interfaces.
Furthermore we assume that the wire is so short that we can neglect inelastic scattering.
In this case the kinetic equation for
the distribution function inside the wire becomes 
$-D\partial_x^2 F =0$.
After solving this equation with the appropriate boundary conditions
one finally finds the current as
$I= V/(R_l + R_{\rm wire}+ R_r)$ with
\begin{eqnarray}
R_r  &= & 1/( 2e^2 {\cal A} N_0 \Gamma_l ) \\
R_{\rm wire } &=&L/(2 e^2 {\cal A} N_0 D ) \\
R_l  &= & 1/( 2 e^2 {\cal A} N_0 \Gamma_r ) 
\end{eqnarray}
as one expects for three resistors in series.

We now discuss the quantum correction to the
current. 
Following \cite{schwab01} we divide the quantum correction
to the current in two contributions:
The first contribution
has the meaning of a   correction to the conductance, $\delta I^{(1)}= V  \delta G$.
The second one is due to the
quantum correction of the distribution function and can be interpreted as the 
redistribution of the voltage along the interfaces and the wire, $\delta I^{(2)} = G \delta V$. 
Both terms are necessary in order to ensure current conservation.
By exploiting current conservation in the structure, $\delta I_l = \delta I_{\rm wire}=\delta I_r$,
and fixing the voltage drop over the whole system to $V$, so that
the sum $\delta V_l + \delta V_r + \delta V_{\rm wire}$ is zero,  
it is possible to eliminate $\delta I^{(2)}$ from the above equations to get
for  the correction to the current :
\begin{equation} \label{eq11}
\delta I = { R_l \delta I_l^{(1)} + R_r \delta I_r^{(1)} +
 R_{\rm wire } \delta I_{\rm wire }^{(1)} \over R_l + R_r + R_{\rm wire}}.
\end{equation}
To proceed further we need the explicit form of $\delta I^{(1)}_{l,r, {\rm wire}}$.
For an interface  attached to an ideal lead on one side
the quantum correction to the current is controlled by the correction to the
density of states on the other side,
\begin{eqnarray}
\delta I_l^{(1)}  &= &  e {\cal A}\Gamma_l \int \D \epsilon  
  \delta N (\epsilon , 0) [ F_\epsilon(0)- F_\epsilon^l ] \label{tunncurrleft}\\
\delta I_r^{(1)}  & = &  e  {\cal A} \Gamma_r \int  \D \epsilon 
  \delta N (\epsilon , L) [ F_\epsilon^r- F_\epsilon(L) ].\label{tunncurrright}
\end{eqnarray}
The limit where both sides of the interface are in thermal equilibrium has been studied many times 
in the literature \cite{altshuler84,nazarov89}. 
Out of equilibrium we obtain the density of states correction as 
\begin{eqnarray}
\delta N (\epsilon, { x} ) &= &
-N_0 \int { \D \omega \over 2 \pi}
S({ x},{ x}) \\
S({ x},{ x}) &= & \Im  \int \D  { x}_1 \cr
 & & \times  F_{\epsilon -\omega}({ x}_1) \rho_\omega({ x},{ x}_1)
             \Phi_\omega({ x}_1, { x})
,\end{eqnarray}
where $\rho_\omega({ x},{ x}_1)$ describes the spreading of a charge injected into the system at 
${ x}_1$; 
it satisfies the equation
\begin{equation} \label{eq16}
( - \I \omega  - D\partial_{ x}^2 ) \rho_{\omega}({ x},{ x}')  = 
 e \delta({ x}-{ x}') 
.\end{equation}
The quantity $\Phi_\omega({ x}_1,{ x})$ is the electrical potential at ${ x}_1$ 
of a charge that has been injected at ${ x}$.
It is given by the product 
of the dynamically screened Coulomb interaction with the diffusion propagator
\begin{equation}
e^2 \Phi_{\omega}({ x},{ x}')  =  \int \D { x}_1 V_{\omega}( { x} ,{ x}_1) 
    \rho_{\omega}({ x}_1,{ x}'). 
\end{equation}
$\rho$ and $\Phi$ depend on the details of the device under consideration and we will come back to them 
below.

The expression for the correction to the current in the wire has been obtained diagrammatically
in \cite{schwab01} and is given by
\begin{eqnarray}
\label{wirecurr}
\delta I_{\rm wire}^{(1)}(x)  &=& - e D N_0 {\cal A} \int \D \epsilon \int {\D \omega \over 2 \pi}
\partial_x[ F_\epsilon(x) S(x, x)] \\
& &+ 2 e D N_0 {\cal A} \int \D \epsilon \int {\D \omega \over 2 \pi} 
F_\epsilon( x) \partial_{x_1}S(x,x_1)|_{x_1=x}.
\nonumber
\end{eqnarray}
In general $\delta I_{\rm wire}^{(1)}(x)$ depends on the position $x$.
Eq.~(\ref{eq11}) however is constructed in such a way that 
the spacial average $\delta I_{\rm wire}^{(1)} =L^{-1} \int \D x \delta I_{\rm wire}^{(1)}(x)$ has to be inserted.

Eqs.(\ref{tunncurrleft},\ref{tunncurrright},\ref{wirecurr}) 
allow to calculate the quantum corrections far from thermal equilibrium
for an arbitrary geometry. One observes that the only ingredients are
the position dependent distribution function $F$, the charge density $\rho$, and the field $\Phi$.
We have already discussed the distribution function $F$ including the relevant boundary conditions 
in Eqs.~(\ref{eq6}), (\ref{eq7}) and below.
Inside a diffusive wire the charge density $\rho$ satisfies Eq.~(\ref{eq16}). 
At the boundaries with the left and right reservoir one may derive the matching conditions
\begin{eqnarray}
D \partial_x \rho_\omega(x,x')|_{x=0} &= & \Gamma_l \rho_\omega(0,x') \cr
D \partial_x \rho_\omega(x,x')|_{x=L}  & = & - \Gamma_r \rho_\omega(L,x')
,\end{eqnarray}
to be compared with Eqs.~(\ref{eq6}) and (\ref{eq7}).
A careful analysis is also required for the field $\Phi_\omega({ x},{ x}')$.
In the case of good metallic screening an injected charge is almost instantly screened
 so that the wire will be electrically neutral
with the exception of a thin surface layer. In this case 
one has inside the wire
\begin{equation}
-\sigma \partial_x^2 \Phi_\omega({ x},{ x}') = e \delta({ x}-{ x}')
,\end{equation}
where $\sigma = 2 e^2 D N_0 {\cal A}$ is the conductivity.
In the absence of surface charges the boundary conditions for $\Phi$ and $\rho$ are identical. In the presence
of these charges, however, this is not the case.
%
%
%

Many special cases where the general formalism discussed above applies have been discussed in 
the literature. For example standard Coulomb blockade physics is found when the resistance of
the system is dominated by one of the interfaces \cite{nazarov89,levitov1997}. In this limit a 
non-equilibrium analysis is not necessary since the distribution function has the equilibrium form on both
sides of the interface. 
The limit of two highly resistive interfaces has been studied in \cite{kamenev96,weber01} and
the opposite limit of a resistive wire and no interface barriers has been discussed
 in \cite{nagaev94,schwab01}.
In the following  we concentrate  on a short resistive wire with interfaces assuming 
temperatures of the order of and lower than the
Thouless energy, $\hbar D/L^2$. 

\begin{figure}
\includegraphics[height=4.9cm]{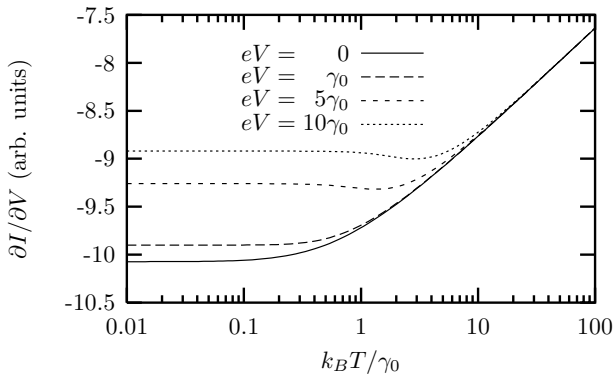}
\caption{\label{Fig1}Temperature dependence of the conductance for different values of the bias voltage. 
$\partial I/ \partial V$ is in units of $e^2/h$ and must be multiplied with
the non-universal number $A$ defined in the Eq.~(\ref{eq23}).
$\gamma_0$ is
the energy of the lowest diffusive mode; in the absence of interface barriers $\gamma_0 = \pi^2 \hbar D/L^2$,
for strongly resistive interfaces $\gamma_0 \to 0$.}
\end{figure}
\begin{figure}
\includegraphics[height=4.9cm]{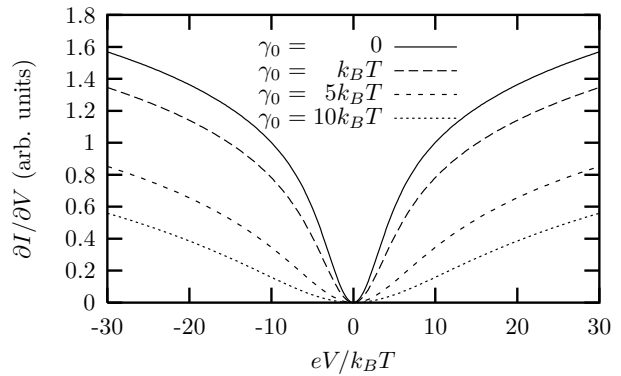}
\caption{\label{Fig2}Voltage dependence of the conductance. 
In all curves the linear conductance has been subtracted.}
\end{figure}

We start by expanding the charge density $\rho$ in diffusive modes, 
\begin{eqnarray}
\label{modes}
\rho_\omega(x,x') &=  &e \sum_n { f_n(x) f_n(x') \over -\I \omega + \gamma_n}  
,\end{eqnarray}
where the (normalized) functions $f_n(x)$ are obtained by the eigenvalue equation
\begin{equation} \label{eq22}
-D \partial_x^2 f_n(x) = \gamma_n f_n(x)
.\end{equation}
In the zero dimensional limit we approximate the sum in Eq.~(\ref{modes}) by retaining only
the eigenmode with the lowest energy, i.e.\
$\rho_\omega(x,x') \to e f_0(x) f_0(x')/(-\I \omega + \gamma_0)$.
This approximation is justified when the energy scales related to the 
temperature and to the voltage remain below the energy of the second lowest diffusive mode.
Let us for the moment ignore charging effects. Then the field $\Phi_\omega(x,x')$ is frequency independent
and one observes that the frequency dependent factors in all the contributions
to the current are identical,
$\delta I(T,V) \sim  F^l_{\epsilon - \omega} F^r_{\epsilon}/(-\I \omega + \gamma_0)$.
The explicit result reads
\begin{eqnarray} \label{eq23}
\delta I &= &
- A  {e \over 2 \pi } \int_0^\infty \! \! \D \eta  \E^{-\gamma_0 \eta}  
\left[{\pi T \over \sinh( \pi T \eta )} \right]^2 \! \sin( e V \eta ) 
\end{eqnarray}
where only the dimensionless number $A$ and the quantity $\gamma_0$ depend on the 
details of the system under consideration.
Notice that the integral has to be cut off at short times in order to avoid a logarithmic divergence.
Let us first  discuss the temperature and voltage dependence of $\delta I$, 
before determining $A$ and $\gamma_0$ explicitly in the two limits of
perfectly transparent interfaces and for interfaces with low transparency.
Figure \ref{Fig1} shows $\partial I /\partial V/(A e^2/2\pi)$ as a function of temperature,
the classical conductance has been subtracted.
At high temperature there is a logarithmic behavior, which saturates below $T_{\rm sat} \sim \max(\gamma_0, eV)$.
Figure \ref{Fig2} shows the voltage dependence of the conductance. Note that the linear 
conductance has been subtracted.
For $\gamma_0 \ll T$ the conductance scales with voltage over temperature, while  when $\gamma_0$ is large the
relevant scale for conductance variations is $\gamma_0$.

How large are the amplitude $A$ and the energy of the lowest diffusion mode $\gamma_0$?
In the case of two well transmitting interfaces, $G_{\rm wire} \ll G_l, G_r$, 
the eigenfunction of Eq.~(\ref{eq22}) with the lowest eigenvalue is
\begin{equation}
f_0(x) =  \sqrt{ 2 \over L } \sin( \pi x  /L), \gamma_0 = \pi^2 D/L^2. 
\end{equation}
The distribution function and the potential $\Phi$ are determined as
\begin{eqnarray}
F(x) &  =  & [(L-x) F_l + x F_r] /L \\
\Phi_\omega(x,x') & = & {e \over G_{\rm wire} } 
\left\{ \begin{array}{lll}
(L-x')x/L^2 & & x< x' \\
(L-x) x'/L^2 & & x> x'
\end{array} \right. 
\end{eqnarray}
and the amplitude of the correction to the current is found to be $A =64/\pi^4 -4/\pi^2   \approx 0.25$.
In the opposite limit, $G_{\rm wire} \gg G_l, G_r$ the eigenvalue equation (\ref{eq22})
may be solved perturbatively in the barrier transparency and one obtains
\begin{eqnarray}
f_0(x) & = & 1/\sqrt{L}, \gamma_0 = (\Gamma_l + \Gamma_r )/L \\ 
F(x)  &  = & ( \Gamma_l F_l + \Gamma_r F_r )/(\Gamma_l + \Gamma_r) \label{eq29} \\
\Phi_\omega(x,x')  & = & {e/( G_l + G_r) } \label{eq30}
,\end{eqnarray}
which leads to $A = 2 \Gamma_l \Gamma_r /( \Gamma_l + \Gamma_r )^2$.

It is useful at this point to  briefly discuss how the charging effects modify the above results.
In the case of highly transmitting interfaces, we assume
that the accumulated charge is proportional to the field as
$\rho(x) = C \Phi(x)$, where $C$ is a capacitance per unit length. This leads to a diffusion equation 
for the field
\begin{equation}
-\I \omega C  \Phi_\omega(x,x') - \sigma \partial_x^2 \Phi_\omega(x,x') = e \delta(x-x')
\end{equation}
with the solution
\begin{equation}
\Phi_\omega(x,x') =  2 e \sum_n {\sin(n \pi x / L) \sin( n \pi x'/L) \over
  - \I \omega (C L ) + (\pi n )^2 G_{\rm wire}}
,\end{equation}
and the correction to the current is modified according to
\begin{eqnarray}
\delta I  &=  &- { e \over 2 \pi }
\int_0^\infty \D \eta \E^{-\eta \gamma_0 } 
\left[ { \pi  T \over \sinh ( \pi T \eta ) } \right]^2 \sin( e V \eta) \cr
&&\times \sum_n A_n\{1- \exp[-\eta (\pi n)^2/R(CL)] \}  \label{eq33}
.\end{eqnarray}
The numbers $A_n$ depend on the wave-function of the diffusive mode $n$ and $R=G_{\rm wire}^{-1}$.
Notice that although the charge density $\rho_\omega(x,x')$ is zero dimensional
many diffusive modes have to be taken into account in the field $\phi_\omega(x,x')$.
For a small capacitance $C$ the charging simply cures the short time divergence in the integral in Eq.~({\ref{eq23}}).
For a larger capacitance the $I$-$V$-characteristics is no longer a universal function.
For the system with poorly transmitting interfaces, the field in Eq.~(\ref{eq30}) 
has to be replaced by
$\Phi_\omega(x,x') = {e / (-\I \omega \tilde C + G_l + G_r ) }$, where we denoted the capacitance of 
the system by $\tilde C$.
The modification to the current is analogous to Eq.~(\ref{eq33}).

In several respects our results agree with \cite{weber01} and \cite{golubev2001}.
We find, in the case of charge neutrality, 
a universal $I$-$V$-characteristics and, over a certain range of temperature,
a logarithmic correction to the conductance.
In some points, however, our results are remarkably different from \cite{weber01,golubev2001}.
For the closed system our improved treatment of the ``zero mode'' allows a more precise calculation
of the amplitude of the $\ln T$ behaviour in the conductance than \cite{weber01}.
The major difference however concerns
the low temperature saturation of $\partial I/ \partial V$. We find that the scale for this saturation 
is set by the energy of the lowest diffusive mode in the system. 
This scale seems to be absent in \cite{golubev2001}, and the origin of this discrepancy is not clear to us.

Finally, as far as the experiments are concerned, a $\ln T$ behavior in the linear resistivity of 
a short metallic bridge together
with an $I$-$V$-characteristics which agrees well with the universal function 
(\ref{eq23}) has been observed in \cite{weber01}.
In \cite{golubev2001} it has been suggested that the effect might be due to the Coulomb effects 
in a diffusive wire,
whereas in \cite{weber01} the Coulomb correction to the tunneling conductance has been suggested 
as the explanation. 
Our work shows that also in the intermediate regime with both diffusive and interface resistivity
the predicted $I$-$V$-characteristics does not change and thus agrees with the experimentally observed one.
Furthermore we can rule out a purely diffusive conductor:
In \cite{weber01} the Thouless energy, which sets the scale for the lowest diffusive mode in an 
open system and 
therefore the low temperature saturation of the conductance,
has been estimated to be of the order of several Kelvin,
whereas the $\ln T$ is observed down to $100$mK.
In the case with resistive interfaces on the other hand the energy of the lowest diffusive mode
is reduced,
$ \gamma_0  \sim  \hbar D/L^2 (R_{\rm wire} / R) \ll \hbar D/L^2$. From this consideration we conclude that
in the experiment the diffusive resistance is considerably smaller than the interface resistance.
A further hint for the importance of interfaces is found from the prefactor $A$:
For the open system we found $A\approx 0.25$ and $A= 2 G_l G_r/(G_r+G_r)^2$ in the tunnel limit.
The experimental values \cite{weber01} are between $A\approx 0.43 \dots 0.7$, i.e.\ closer to the 
the tunnel limit than to the open system. 
In order to check these ideas it would be of interest to modify experimentally the resistance
of the interface relative to the short bridge and observe both a change
of the prefactor of the $\ln T$ behavior and of the saturation temperature.

In conclusion, 
we calculated the Coulomb interaction contribution to the current through structures which are composed of
diffusive pieces and resistive interfaces.
Our general formalism  agrees with earlier studies on the Coulomb correction 
to the tunneling conductance \cite{altshuler84,nazarov89,levitov1997} 
and on the Coulomb correction in diffusive conductors \cite{altshuler85,altshuler1980}.
In contrast to those earlier studies our formalism treats both effects on equal footing and is valid even
far from thermal equilibrium.
We concentrated on the zero dimensional limit valid for temperatures
below the Thouless energy and shown that our theoretical results provide an explanation
of the experimental findings of \cite{weber01}.

The authors would like to acknowledge discussions with U.~Eckern, J.~Kroha, E.~Scheer, 
and H.~Weber on the subject.
This work was supported from the
German-Italian exchange program Vigoni 2002 (CRUI and DAAD), 
the Deutsche Forschungsgemeinschaft through SFB484 and
E.U. by Grant RTN 1-1999-00406.

\end{document}